\documentstyle[preprint,prb,aps,eqsecnum,psfig]{revtex}
\tighten

\begin{document}
\draft

\title{Exact Eigenstates of Tight-Binding Hamiltonians on the Penrose Tiling}
\author{Przemys{\l}aw Repetowicz, Uwe Grimm, and Michael Schreiber}
\address{Institut f\"ur Physik, 
         Technische Universit\"at Chemnitz, 
         D-09107 Chemnitz, Germany}
\date{\today}
\maketitle

\begin{abstract}%
We investigate exact eigenstates of tight-binding models on the planar
rhombic Penrose tiling. We consider a vertex model with hopping along
the edges and the diagonals of the rhombi. For the wave functions, we
employ an ansatz, first introduced by Sutherland, which is based on
the arrow decoration that encodes the matching rules of the
tiling. Exact eigenstates are constructed for particular values of the
hopping parameters and the eigenenergy. By a generalized ansatz that
exploits the inflation symmetry of the tiling, we show that the
corresponding eigenenergies are infinitely degenerate. Generalizations
and applications to other systems are outlined.
\end{abstract}

\pacs{PACS numbers:
 71.23.Ft,  %Electronic structure: Quasicrystals%
 05.60.+w,  %Transport processes: theory%
 71.23.An,  %Theories and models; localized states%
 71.30.+h   %Metal-insulator transitions and other electronic transitions%
}

\narrowtext

\section{Introduction}
\label{sec1}

The discovery of quasicrystals by Shechtman et al.\ \cite{Shechtman}
stimulated wide interest in the physics of these materials which are
intermediate between periodic and random structures.  Besides
icosahedral quasicrystals, which are aperiodic in all three dimensions
of space, also periodically layered structures with planar
quasiperiodic order and non-crystallographic rotational symmetries
were found, comprising dodecagonal,\cite{INF}
decagonal,\cite{Bendersky} and octagonal \cite{WCK} phases with
twelve-, ten-, and eightfold symmetry, respectively.  Although the
fundamental question ``where are the atoms?'', raised e.g.\ in
Ref.~\onlinecite{Bak}, has only been answered partially to date, most
structure models of quasicrystals are based on two- or
three-dimensional quasiperiodic tilings or their disordered versions
(random tilings).

An important and exciting problem in condensed-matter physics is
whether the quasiperiodic structure leads to new and unexpected
physical properties. In particular transport properties, as for
instance electric or heat conductance, are strongly effected by the
non-periodic order.  Indeed, many quasicrystalline alloys are
characterized by very high values of electric resistivity, by a
negative temperature coefficient of resistivity, and by a low
electronic contribution to the specific heat which points to a small
density of states at the Fermi energy. It is difficult to explain
these striking features, because a rigorous theory of the electronic
structure of quasiperiodic materials does not exist.  For want of a
simple analogy of Bloch theory for quasicrystals, one either carries
out numerical calculations for as large clusters as possible, or one
tries to make exact statements about the electronic wave functions in
simple models.

\begin{figure}
 \centerline{\psfig{figure=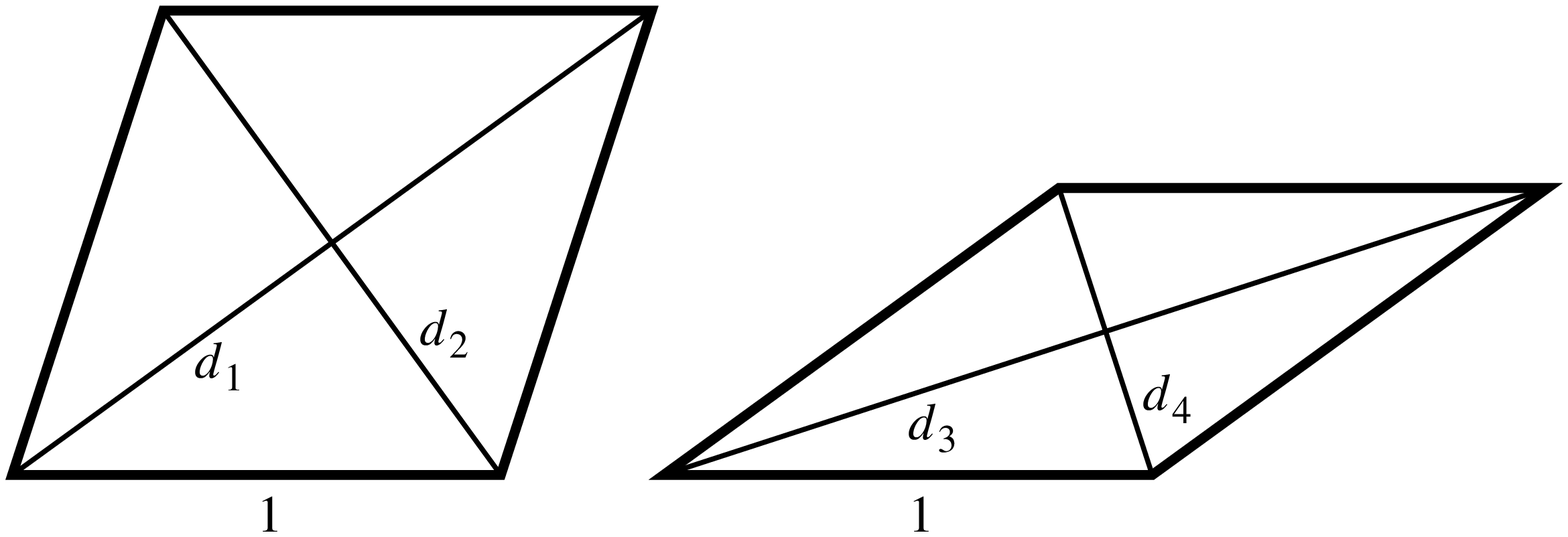,width=\columnwidth}}
 \caption{The two types of rhombi in the Penrose tiling and the
 assignment of hopping integrals $d_{1}$, $d_{2}$, $d_{3}$, and
 $d_{4}$ to their diagonals. The hopping integral along the edges is
 chosen as~1.\label{fig:rhombi}}
\end{figure}

We consider tight-binding models on the two-dimensional rhombic
Penrose tiling.\cite{Penrose} Our models are so-called vertex models
because we locate the atoms at the vertices of the
tiling. Interactions are taken into account only between neighboring
vertices connected by edges or by diagonals of the rhombi. In our
calculations, we restrict ourselves to a single $s$-type atomic
orbital per vertex. This makes the transfer integrals $t_{ij}$
independent of the angular orientation and leads to the following
Hamiltonian
\begin{equation}
H \; = \; \sum_i |i\rangle \,\varepsilon_i\, \langle i| \; + 
\; \sum\limits_{i,j} |i\rangle \, t_{ij}\, \langle j|
\label{hamiltonian}
\end{equation}
where $|i\rangle$ denotes a Wannier state localized at vertex $i$, and
$\varepsilon_i$ are on-site energies. For the hopping integrals
$t_{ij}$, we choose five different values $1$, $d_1$, $d_2$, $d_3$,
$d_4$, depending on the distance of the vertices $i$ and $j$, see
Fig.~\ref{fig:rhombi}. Here, $t_{ij}=1$ for vertices connected by an
edge of the tiling, $t_{ij}=d_1$ ($d_2$) for the long (short) diagonal
of the `fat' rhombus, and $t_{ij}=d_3$ ($d_4$) for the long (short)
diagonal of the `thin' rhombus, respectively.

As the Penrose tiling is arguably the most popular among the
quasiperiodic tilings, it is not surprising that tight-binding models
defined on the Penrose tiling have been investigated rather
thoroughly. Besides the vertex
model,\cite{Choy,ON,KS,Suth,Oda,KA,Koh,ATFK,LM,YYZY,NBW,RS1,RS2} the
so-called center model was considered,\cite{TFUT1,ATF,FATK,TFA,TFUT2}
where atoms are located in the center of the rhombi, and hopping may
occur between adjacent tiles --- this is nothing but a vertex model on
the dual graph of the Penrose tiling. However, most results rely on
numerical approaches, and only few exact results on the spectrum of
the tight-binding Hamiltonian are known. In particular, so-called
`confined states' have been investigated in detail, both for the
vertex \cite{KS,ATFK} and the center model.\cite{FATK} These are
infinitely degenerate, strictly localized eigenstates corresponding to
a particular value of the energy, which occur as a consequence of the
{\em local}\/ topology of the tiling, see also
Ref.~\onlinecite{RS1}. Furthermore, for a Hamiltonian
(\ref{hamiltonian}) with particular on-site energies $\varepsilon_i$
chosen according to the vertex type at site $i$, the exact
self-similar ground state could be constructed.\cite{Suth} Based on
the same idea, several non-normalizable eigenstates of the center
model and their multifractal properties were obtained
exactly.\cite{TFA} These solutions, restricted to special values of
the hopping integrals, were derived from a suitable ansatz for the
eigenfunctions. According to this ansatz, the wave function at a site
depends only on its neighborhood and on a certain integer number
associated to the site, a `potential', which is derived from the
matching rules of the Penrose tiling.\cite{Suth}

In this paper, we apply the same ansatz to the vertex model on the
Penrose tiling. The solution is more complicated than for the center
model, where the coordination number (i.e., the number of neighbors)
is always equal to four, whereas for the vertex model it varies
between three and seven, or between six and fourteen if we include
neighbors along diagonals, respectively.  For suitably chosen transfer
integrals, we derive exact eigenstates of the Hamiltonian
(\ref{hamiltonian}) and analyze their multifractal behavior. As
observed for the center model,\cite{TFA} we find that these states are
infinitely degenerate, i.e., for fixed value of the energy the
eigenfunctions still involve one free parameter. In order to show
this, we need to generalize the ansatz exploiting the inflation
symmetry of the Penrose tiling.

Our paper is organized as follows. In the subsequent section, we
discuss the labeling of the rhombi with two kind of arrows and the
associated potentials.  In Sec.~\ref{sec3}, we introduce the ansatz
for the wave function and solve the tight-binding equations for two
cases, the first one with $\varepsilon_i = 0$ and $d_i \ne 0$, and the
second one with $d_i=0$ but various on-site energies. A generalized
ansatz, based on the inflation symmetry of the tiling, is considered
in Sec.~\ref{sec4}. In Sec.~\ref{sec5}, we perform a fractal analysis
of the wave functions, i.e., we calculate the generalized dimensions.
Finally, we conclude in Sec.~\ref{sec6}.

\section{Edge-labeling and Potentials}
\label{sec2}

\begin{figure}
 \centerline{\psfig{figure=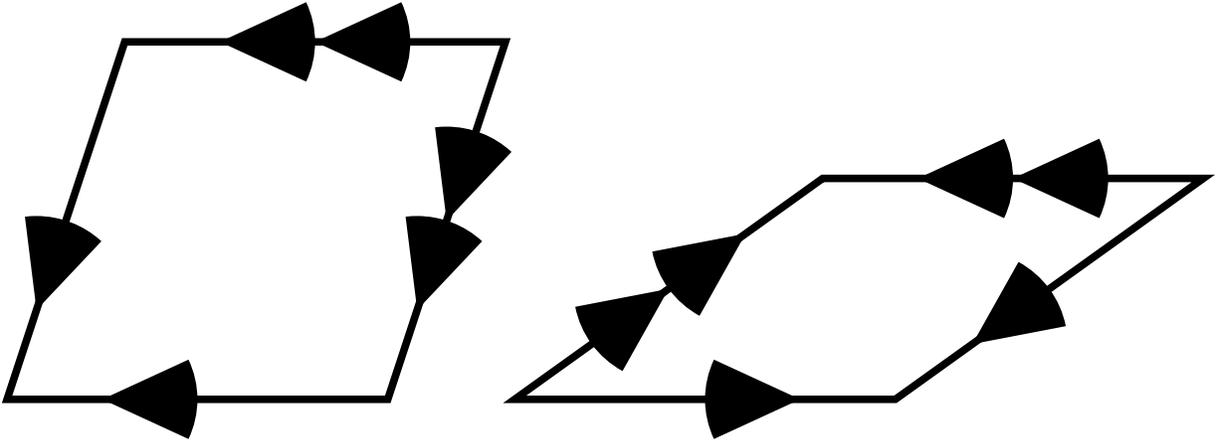,width=\columnwidth}}
 \caption{Arrow decoration of the Penrose rhombi. Note that our
          decoration differs from that used in 
          Refs.~\protect\onlinecite{Suth} and
          \protect\onlinecite{TFA} in the direction of the 
          arrows.\label{fig:arrowedrhombi}}
\end{figure}

\begin{figure}
 \centerline{\psfig{figure=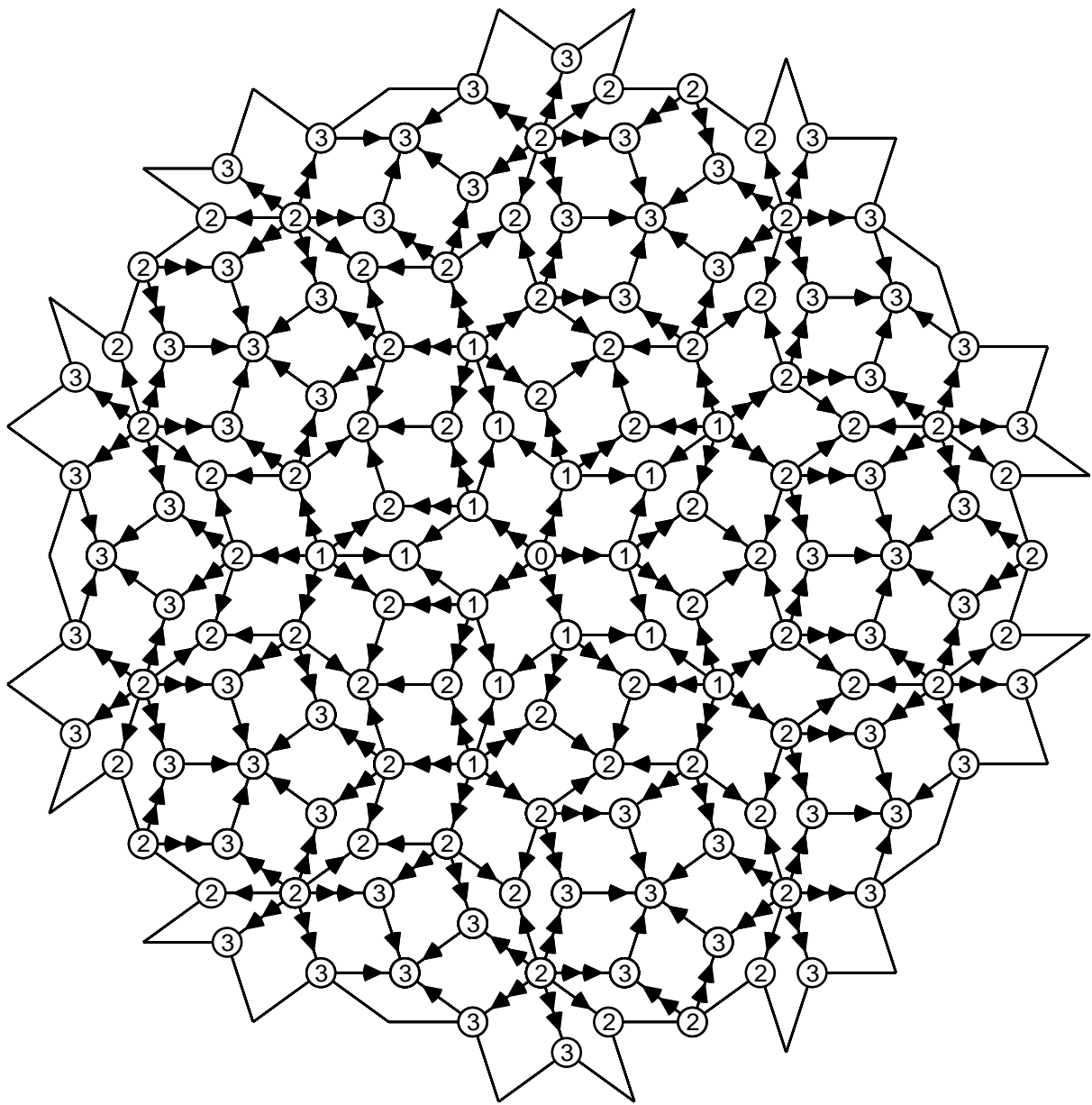,width=\columnwidth}} 
\caption{The double-arrow potential for a patch of the 
         Penrose tiling.\label{fig:pot1}}
\end{figure}

Following de Bruijn,\cite{deBruijn} we mark the rhombi with single and
double arrows as shown in Fig.~\ref{fig:arrowedrhombi}. The matching
rules require that arrows on adjacent edges match.  Fixing a certain
site $O$ as the origin, we assign to a site $i$ two integers $n(i)$
and $m(i)$ which count the number of single and double arrows,
respectively, along an arbitrary path connecting the origin $O$ and
site $i$.  This is well-defined because, along any closed path, the
total number of single and double arrows vanishes, as can be seen from
Fig.~\ref{fig:arrowedrhombi}. We refer to these integers $n(i)$ and
$m(i)$ as `potentials' at site $i$ because they are integrals of the
two vector fields defined by the arrows.  The distributions of the
potentials are rather irregular and show the following properties:
\begin{itemize}
\item The single-arrow potential $n(i)$ is directly related to the sum
      $t(i)\in\{1,2,3,4\}$ of the five-dimensional indices denoting
      the translation class of the site $i$. It takes only two values:
      $n(i)=0$ if $t(i)\in\{2,3\}$ and $n(i)=1$ if $t(i)\in\{1,4\}$
      (provided the origin has translation class $t(O)\in\{2,3\}$).
\item The double-arrow potential $m(i)$ is unbounded. Its distribution
      on a finite patch is shown in Fig.~\ref{fig:pot1}. For a detailed 
      discussion of the distribution, see Ref.~\onlinecite{Suth}.
\end{itemize}
The potential $m(i)$ is the key ingredient in the construction of
exact eigenfunctions of tight-binding Hamiltonians on the Penrose
tiling.

\section{Solutions of the tight-binding equations}
\label{sec3}

We want to construct solutions of the tight-binding equations
\begin{equation}
\sum_j t_{ij}\,\phi_{j}\; = \;  (E-\varepsilon_{i})\: \phi_i \; ,
 \label{eq:tight_binding_equations}
\end{equation}
where we sum over all neighbors $j$ of the site $i$. As an ansatz, we
demand that the wave function amplitude $\phi_{i}$ at site $i$ depends
solely on the vertex type $\nu(i)\in\{1,2,\ldots,8\}$ and on the
potential $m(i)$. This leads to the following ansatz
\begin{equation}
 \phi_i \; = \; A_{\nu(i)} \,  \beta^{\, m(i)} \; .
 \label{eq:ansatz}
\end{equation}
The eight vertex types of the Penrose tiling are shown in the top row
of Fig.~\ref{fig:vertex_types}.  The corresponding eight amplitudes
$A_{\nu}$ and $\beta$ are parameters.

\subsection{The case \protect\boldmath{$\varepsilon_i = 0$}}

\begin{figure}
 \centerline{\psfig{figure=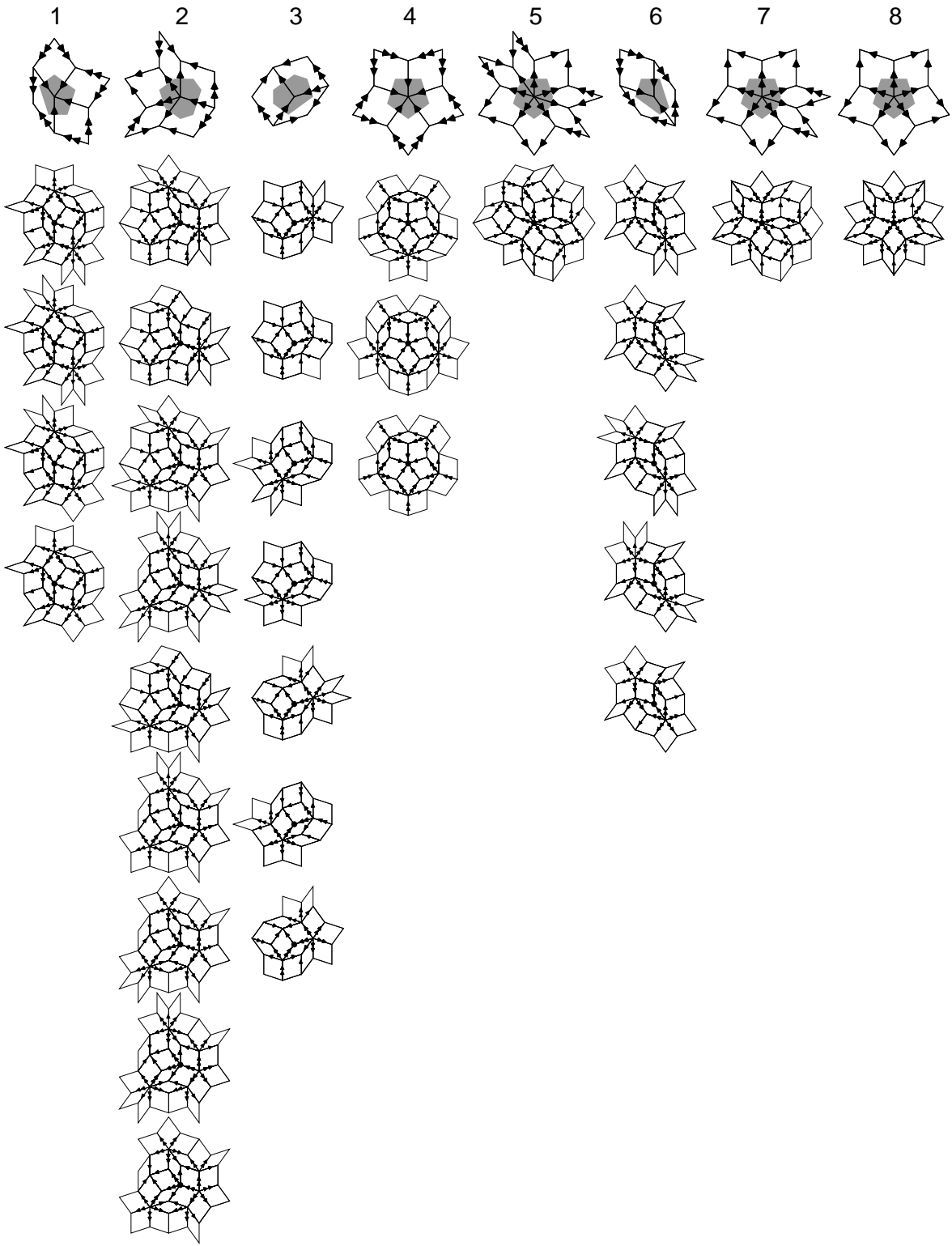,height=0.94\textheight}} \caption{The
 eight vertex types of the Penrose tiling (top row) with the
 corresponding Voronoi cells (shaded), and the corresponding
 second-order vertex types (below).\label{fig:vertex_types}}
\end{figure}

\begin{figure}
 \centerline{\psfig{figure=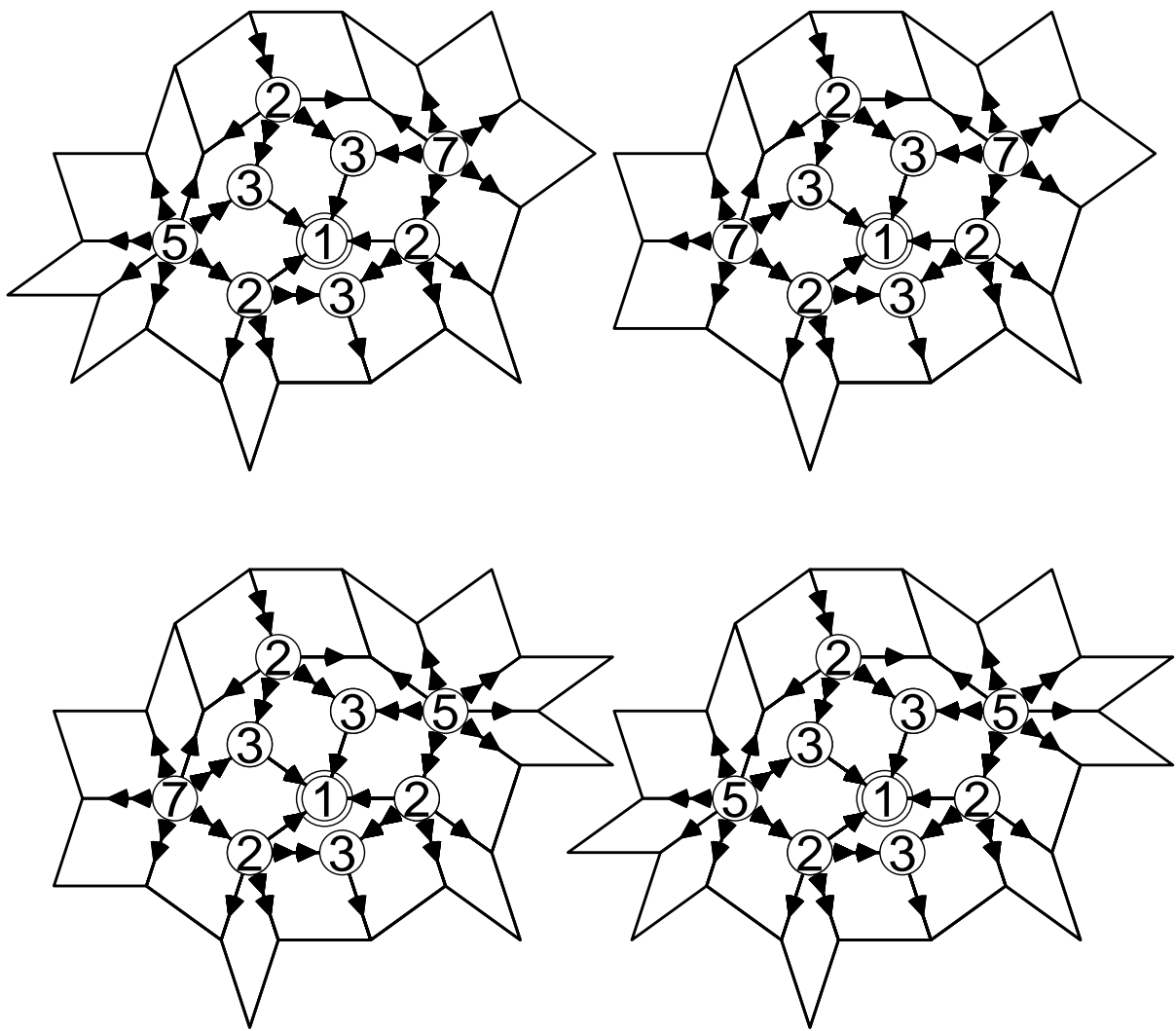,width=\columnwidth}}
\caption{Second-order vertex types corresponding to a central vertex
         of type 1. Here, the encircled numbers denote the vertex
         types, not the potential.\label{fig:vertex_types21}}
\end{figure}

For simplicity, we first concentrate on the case with on-site energies
$\varepsilon_i=0$. With the ansatz (\ref{eq:ansatz}), the infinite set
of equations (\ref{eq:tight_binding_equations}) reduces to a finite
set comprising as many equations as there are second-order vertex
types in the tiling. By a second-order vertex type we mean the
neighborhood of a site up to its second coordination zone. There are
31 different second-order vertex types in the Penrose tiling, these
are shown in Fig.~\ref{fig:vertex_types}, grouped together according
to the first-order vertex type of the central site given in the top
row. Thus, we have $31$ linear equations in the 14 variables $A_\nu$
($\nu=1,\ldots,8$), $\beta$, $d_1$, $d_2$, $d_3$, $d_4$, and $E$. As
it is straightforward to derive the equations from the second-order
vertex types of Fig.~\ref{fig:vertex_types}, we refrain from listing
them here.  Instead, we consider as an example only the second-order
vertex types in the first column of Fig.~\ref{fig:vertex_types}, which
we show again in Fig.~\ref{fig:vertex_types21} (rotated by 90 degrees)
together with the corresponding values of the potential $m(i)$. This
yields the following four equations
\begin{eqnarray}
E A_1
& = & d_4 A_3\beta + 2(A_2 + A_3) + d_1 (A_2 + A_5 + A_7) \beta^{-1}
 \nonumber \\ 
& = & d_4 A_3\beta + 2(A_2 + A_3) + d_1 (A_2 + A_5 + A_5) \beta^{-1}
\nonumber \\ 
& = & d_4 A_3\beta + 2(A_2 + A_3) + d_1 (A_2 + A_7 + A_5) \beta^{-1}
\nonumber \\ 
& = & d_4 A_3\beta + 2(A_2 + A_3) + d_1 (A_2 + A_7 + A_7) \beta^{-1}
\nonumber \\ 
\label{eq:system_of_equations}
\end{eqnarray}
two of which (the first and the third) are identical because the
corresponding patterns are mirror images of each other.

At first sight, as the number of variables, $14$, is much smaller than
the number of equations, $31$, one might expect that only the trivial
solution ($\phi_i\equiv 0$) exists.  However, this is not the case,
for suitably chosen values of the hopping parameters $d_1$, $d_2$,
$d_3$, $d_4$, and the energy $E$, non-trivial solutions exist, because
the equations are not independent. To see this, note that the
second-order vertex types within one column of
Fig.~\ref{fig:vertex_types} differ only slightly from each other,
which means that the corresponding equations are also very similar as
can be seen in the example (\ref{eq:system_of_equations}). Thus, they
can be substantially simplified by subtraction. For example, the
differences between the equations in (\ref{eq:system_of_equations})
result in the single equation
\begin{equation}
 d_1 (A_5 -  A_7) \;=\; 0 
\end{equation}
which implies $A_5=A_7$ (unless $d_1$ vanishes). {}From the analogous
equations for the other vertex types, it turns out that the amplitudes
$A_{\nu(i)}$ depend only on the translation class $t(i)$ of the site
$i$, rather than on its specific vertex type $\nu(i)$. This means
\begin{eqnarray}
& & A_1 \,=\, A_4 \,=\, A_6 \qquad (t\in\{1,4\}) \nonumber\\ 
& & A_2 \,=\, A_3 \,=\, A_5 \,=\, A_7 \,=\, A_8 \qquad (t\in\{2,3\}) \; .
\label{eq:ampli}
\end{eqnarray}
With this, all equations corresponding to second-order vertex types
with the same central vertex reduce to a single equation, and one is
left with the following eight equations
\begin{eqnarray}
E A_1
& = & d_4 A_2\beta  + 4 A_2 + 3 d_1 A_2 \beta^{-1} \nonumber \\
& = & 5 A_2 + 5 d_1 A_2 \beta^{-1} \nonumber \\
& = & 2 d_4 A_2 \beta + 3 A_2 + d_1 A_2 \beta^{-1} \nonumber \\
E A_2
& = & (d_1 A_1 + 2 A_2)\beta + 2 A_1  + 2 (d_2+d_3) A_2 + A_2\beta^{-1}
\nonumber \\
& = &  A_1 + 2 d_2 A_2 + (2 A_2 + d_4 A_1)\beta^{-1} \nonumber \\
& = & (3 d_1 A_1 + 5 A_2)\beta + 2 A_1 + 4 d_3 A_2 \nonumber \\
& = & (4 d_1 A_1 + 5 A_2)\beta + A_1 + 2 d_3 A_2 \nonumber \\
& = & (5 d_1 A_1 + 5 A_2) \beta 
\label{eq:simplf_equations}
\end{eqnarray} 
for central vertices of type $1$, $4$, $6$, and $2$, $3$, $5$, $7$,
$8$, respectively.

For this system of equations, we obtain three sets of non-trivial
solutions, expressed in terms of the parameter $\beta$ which may
be chosen freely. Here, we introduce
\begin{equation}
b_{\pm} \; := \; \frac{1}{\beta\pm\beta^{-1}}
\label{eq:bpm}
\end{equation}
to abbreviate the formulae below.\newline
{\em Solution (1):} The wave function has the form
\begin{equation}
\phi_i \;=\; \left\{\begin{array}{ll}
             (1 - 2\beta^2)\beta^{\, m(i)} & \mbox{for $t(i)\in\{1,4\}$}\\
             \beta^{\, m(i)+1}            & \mbox{for $t(i)\in\{2,3\}$} 
             \end{array}  \right.
\label{eq:solution1}
\end{equation}
for transfer integrals and energy given by
\begin{eqnarray}
& & \makebox[0.45\columnwidth][l]{$d_1 = \frac{1}{2}b_{-}$}
\makebox[0.45\columnwidth][l]{$d_2 = -\frac{3}{4}b_{-}+b_{-}^{-1}$}\nonumber\\
& & \makebox[0.45\columnwidth][l]{$d_3 = -\frac{1}{4}b_{-}+
\frac{1}{2}b_{-}^{-1}$} 
\makebox[0.45\columnwidth][l]{$d_4 = b_{-}$} \nonumber\\
& & \makebox[0.45\columnwidth][l]{$E\, = -\frac{5}{2}b_{-}$} 
\label{eq:parameter1}
\end{eqnarray}
{\em Solution (2):} Here,
\begin{equation}
\phi_i \;=\; \left\{ \begin{array}{ll}
             \beta^{\, m(i)+1} & \mbox{for $t(i)\in\{1,4\}$} \\
             \beta^{\, m(i)}   & \mbox{for $t(i)\in\{2,3\}$} 
             \end{array}  \right.
\label{eq:solution2}
\end{equation}
with
\begin{eqnarray}
& & \makebox[0.45\columnwidth][l]{$d_1 = -b_{+}$} 
\makebox[0.45\columnwidth][l]{$d_2 = \frac{3}{2}b_{+} - 
\frac{1}{2}b_{+}^{-1}$}\nonumber \\
& & \makebox[0.45\columnwidth][l]{$d_3 = -(\frac{1}{2}+\beta^2)b_{+}$}
\makebox[0.45\columnwidth][l]{$d_4 = (1-\beta^{-2})b_{+}$}\nonumber \\
& & \makebox[0.45\columnwidth][l]{$E\, = 5 b_{+}$}\label{eq:parameter2}
\end{eqnarray}
{\em Solution (3):} Finally, 
\begin{equation}
\phi_i \;=\; \left\{ \begin{array}{ll}
             \beta^{\, m(i)}   & \mbox{for $t(i)\in\{1,4\}$} \\
             \beta^{\, m(i)+1} & \mbox{for $t(i)\in\{2,3\}$} 
             \end{array}  \right.
\label{eq:solution3}
\end{equation}
where
\begin{eqnarray}
& & \makebox[0.45\columnwidth][l]{$d_1 = -(\frac{1}{2}+\beta^2)b_{+}$} 
\makebox[0.45\columnwidth][l]{$d_2 = \frac{1}{4}b_{+}-
\frac{1}{2}b_{+}^{-1}$}\nonumber \\
& & \makebox[0.45\columnwidth][l]{$d_3 = -(\frac{1}{4}+\beta^{-2})b_{+}$}
\makebox[0.45\columnwidth][l]{$d_4 = -b_{+}$}\nonumber \\
& & \makebox[0.45\columnwidth][l]{$E\, = \frac{5}{2}b_{+}$}
\label{eq:parameter3}
\end{eqnarray}

\begin{figure}
 \centerline{\psfig{figure=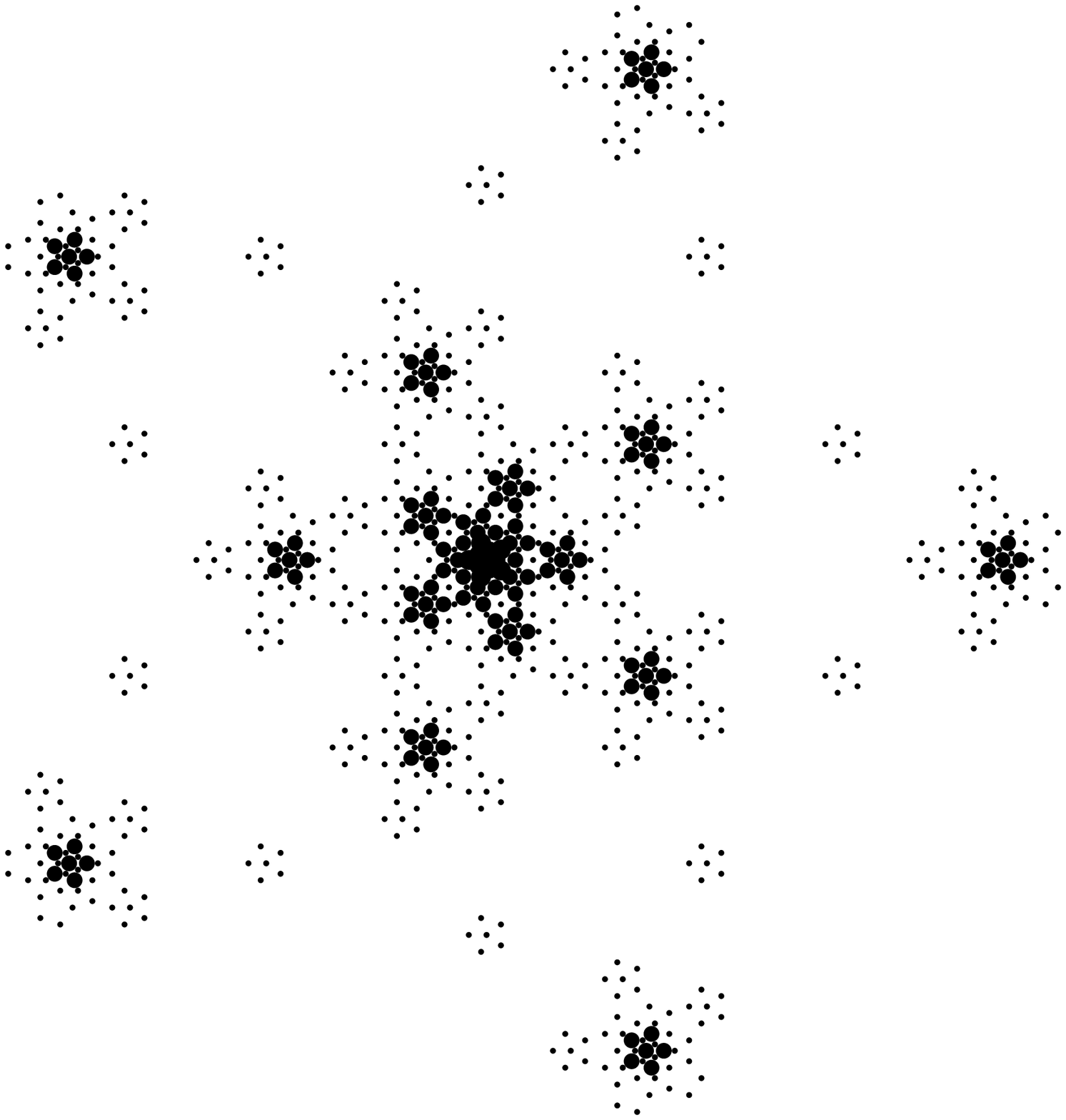,width=0.5\columnwidth} 
             \psfig{figure=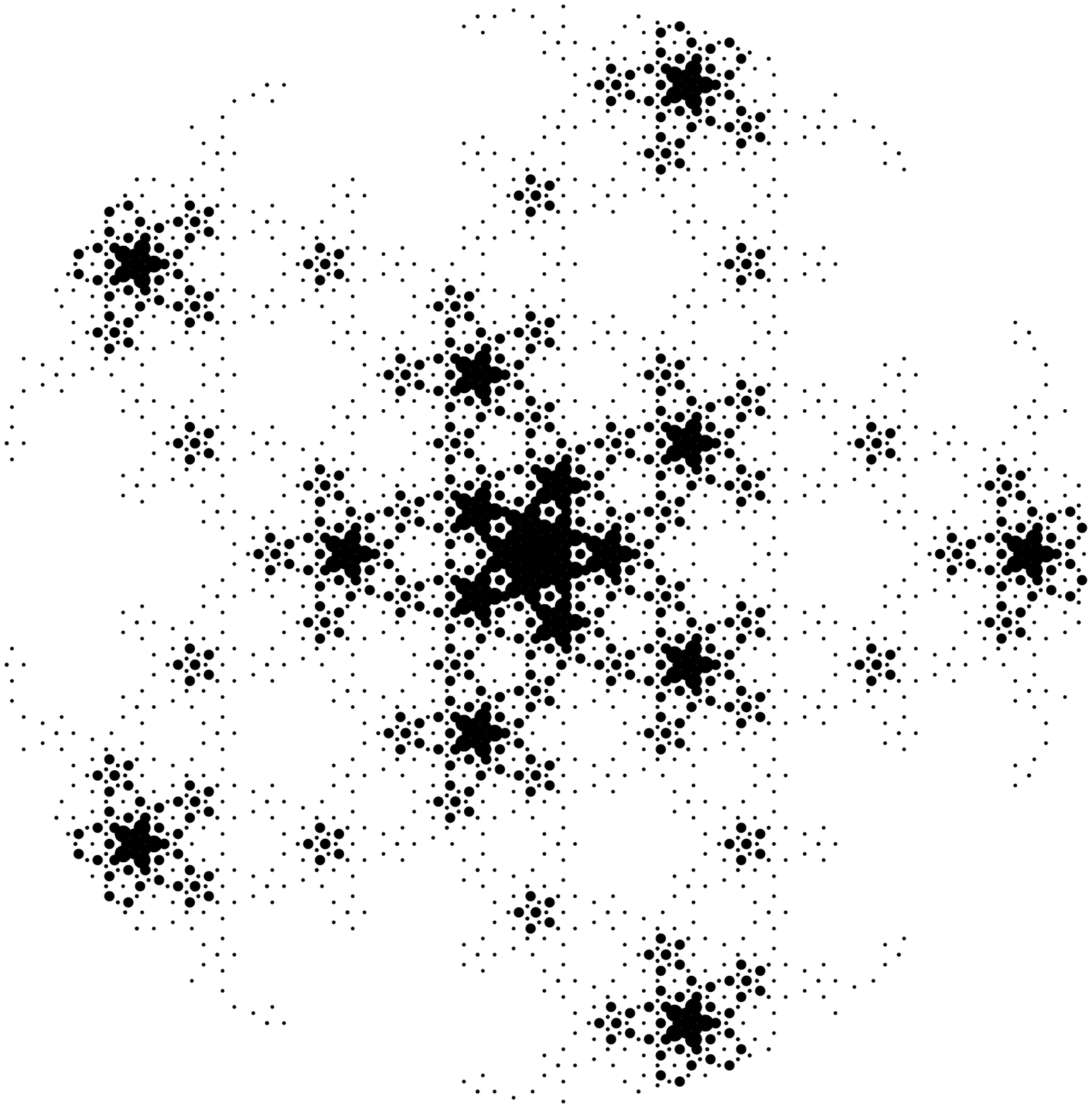,width=0.5\columnwidth}}
 \centerline{\psfig{figure=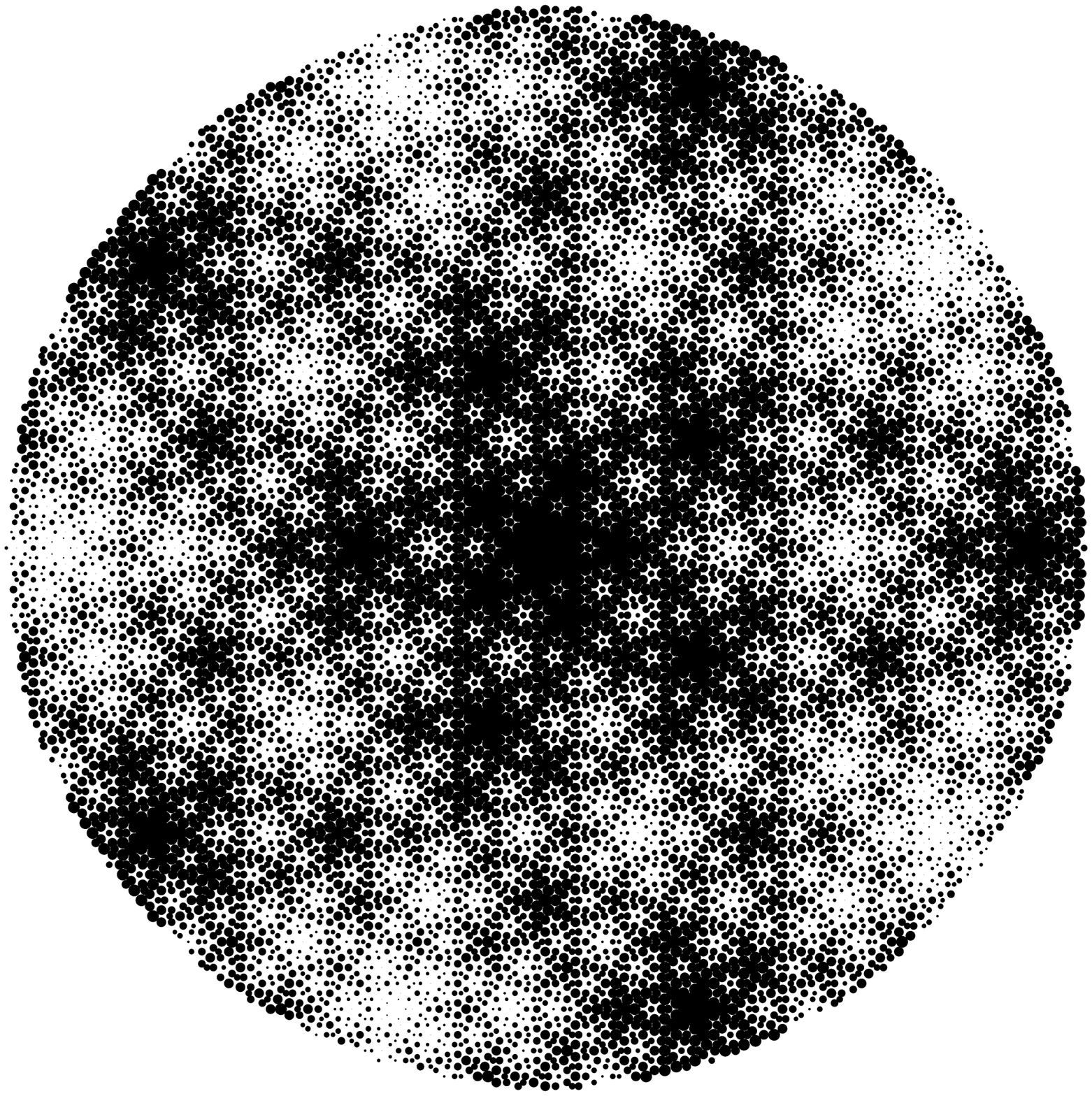,width=0.5\columnwidth} 
             \psfig{figure=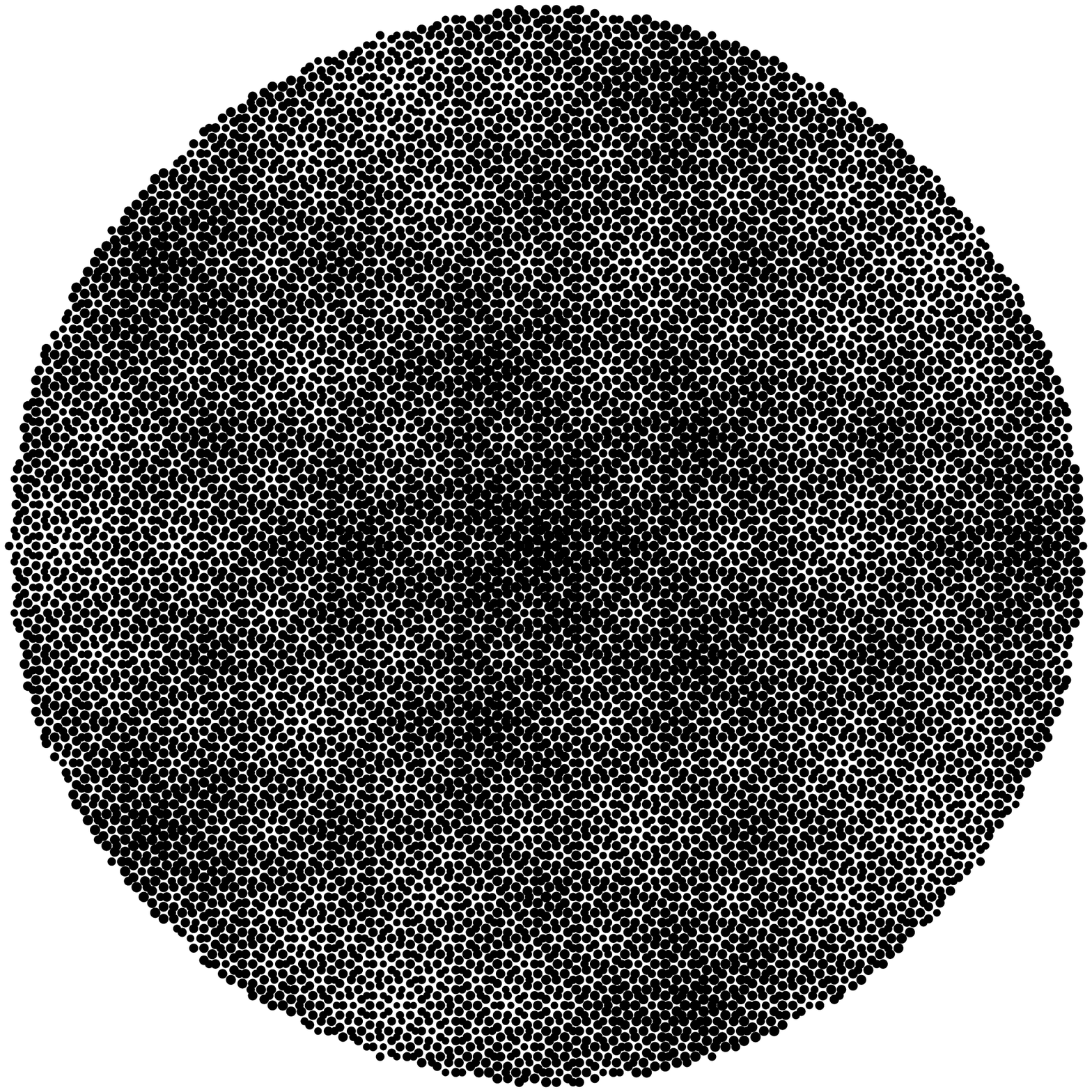,width=0.5\columnwidth}}
\caption{Wave functions (\ref{eq:solution1}) for four different values
         $\beta = 0.1$, $0.2$, $0.6$, and $0.9$. On a finite patch of
         ${\cal N}=16\, 757$ vertices obtained by sevenfold inflation
         of a vertex of type $4$, the wave function has been
         normalized to $\sum_{i} |\phi_i|^2=1$.  The radii of the
         circles encode $|\phi_i|^2$, in units of the edge length they
         have been chosen as $R=0$ for ${\cal N} |\phi|^2<10^{-2}$,
         $R=1.8\log(10^2 {\cal N} |\phi|^2)/\log(10^4)$ for
         $10^{-2}\le{\cal N}|\phi|^2\le 10^2$ and $R=1.8$ for ${\cal
         N}|\phi|^2>10^2$.\label{fig:wavefunctions}}
\end{figure}

For each of these solutions, there exists an additional solution for a
slightly generalized ansatz 
\begin{equation}
\tilde{\phi}_i = \tilde{A}_{\nu(i),t(i)}\: \tilde{\beta}^{\, m(i)}
\label{eq:genansatz}
\end{equation}
that involves the translation class $t(i)$ at site $i$. Note that for
each vertex type there are only two possible values, $t(i)\in\{1,4\}$
for $\nu(i)\in\{1,4,6\}$ and $t(i)\in\{2,3\}$ for
$\nu(i)\in\{2,3,5,7,8\}$, which were not distinguished in our previous
ansatz, see Eq.~(\ref{eq:ampli}). The wave functions differ from the
solutions given in Eqs.~(\ref{eq:solution1}), (\ref{eq:solution2}),
and (\ref{eq:solution3}) only by an alternating sign which depends on
the translation class
\begin{equation}
\tilde{\phi}_i = (-1)^{t(i)} \phi_i
\label{eq:gensol}
\end{equation}
and by a sign change in the parameters, i.e., $\tilde{d_1}=-d_1$,
$\tilde{d_2}=-d_2$, $\tilde{d_3}=-d_3$, $\tilde{d_4}=-d_4$, and
$\tilde{E}=-E$. Note that the two models differing by this sign change
are not trivially related, because the hopping parameter along the
edges of the tiles does not change its sign --- it is always equal to
$1$. These six solutions exhaust all non-trivial solutions in terms of
the ansatz (\ref{eq:genansatz}), but we note that for a given set of
hopping parameters, i.e., for a given Hamiltonian, this yields at most
one single solution.  Some examples of the wave functions
(\ref{eq:solution1}) for different values of $\beta$ are presented in
Fig.~\ref{fig:wavefunctions}.

\subsection{The case \protect\boldmath{$\varepsilon_i \neq 0$}}

In order to obtain the eigenstates described above, we introduced
parameters in the Hamiltonian (\ref{hamiltonian}) and determined them
by requiring that the ansatz (\ref{eq:ansatz}) fulfills the
tight-binding equations (\ref{eq:tight_binding_equations}). In
Eq.~(\ref{hamiltonian}), we already included the possibility of
site-dependent on-site energies $\varepsilon_i$. In the present case,
it is natural to choose the on-site energies $\varepsilon_i$ according
to the vertex type of site $i$.  That is,
$\varepsilon_{i}=\mu_{\nu(i)}$ with eight parameters $\mu_{1}, \ldots,
\mu_{8}$ according to the eight vertex types of the Penrose tiling.

Of course, we can perform the same analysis as above for the more
general problem --- it just amounts to replacing the left-hand side of
the first three lines of Eqs.~(\ref{eq:simplf_equations}) by
$(E-\mu_{\nu})A_{1}$ with $\nu=1,4,6$, and in the remaining five
lines by $(E-\mu_{\nu})A_{2}$ with $\nu=2,3,5,7,8$, respectively.  We
do not show the explicit solution of the full problem because it is
rather lengthy. Although the general solution contains a few free
parameters, for a given Hamiltonian we still find at most one exact
eigenstate.

In order to compare with Sutherland's result,\cite{Suth} we consider
the case without hopping along the diagonals of the rhombi, i.e.,
$d_1=d_2=d_3=d_4=0$. We can express the solutions in terms of the
three parameters $E$, $\beta$, and $\gamma:=A_{2}/A_{1}$:
\begin{eqnarray}
& & \makebox[0.5\columnwidth][l]{$\displaystyle 
\mu_1 = E - 4 \gamma$}
\mu_2 = E - 2\beta - \beta^{-1} - 2\gamma^{-1} \nonumber \\
& & \makebox[0.5\columnwidth][l]{$\displaystyle 
\mu_3 = E - 2\beta^{-1} - \gamma^{-1}$} 
\mu_4 = E - 5 \gamma \nonumber \\
& & \makebox[0.5\columnwidth][l]{$\displaystyle 
\mu_5 = E - 5 \beta - 2\gamma^{-1}$} 
\mu_6 = E - 3\gamma \nonumber \\
& & \makebox[0.5\columnwidth][l]{$\displaystyle 
\mu_7 = E - 5 \beta - \gamma^{-1}$} 
\mu_8 = E - 5\beta 
\end{eqnarray}
Setting $\gamma = 1$ one recovers Sutherland's solution.\cite{Suth}

Taking into account that we have introduced eight parameters
$\mu_{\nu}$ in our the Hamiltonian, it was almost obvious that
solutions exist. It would be, however, more interesting to introduce
additional parameters in the ansatz for the wave function. In this
way, one might perhaps be able to obtain several eigenstates of a
given Hamiltonian and thus come closer to the general solution of our
problem. This is the subject of the following section.

\section{Generalized ansatz for the eigenfunctions}
\label{sec4}

The Penrose tiling possesses a so-called inflation/de\-flation
symmetry.\cite{Penrose,Suth} In an inflation step, the two types of
rhombi are dissected into smaller pieces that again constitute a
rhombic Penrose tiling, but on a smaller scale with all lengths
divided by the golden ratio $\tau=(1+\sqrt{5})/2$. The inverse
procedure, in which tiles are combined to form larger tiles, is known
as deflation.

The idea now is to generalize the ansatz (\ref{eq:ansatz}) for the
wave function by using the vertex types and potentials of the deflated
tiling in addition to those of the original tiling. Even more general,
one may consider a sequence of $n$ tilings obtained by successive
deflation steps, probing the original tiling on larger and larger
length scales. In this way, we assign to each vertex $i$ of the
original tiling a sequence of integers $\{\nu_{k}(i)\}$,
$k=0,1,\ldots,n$, where $\nu_{k}(i)$ specifies the corresponding
vertex type in the $k$-fold deflated tiling, with $k=0$ referring to
the original tiling.  This leads to the following generalized ansatz
for the wave function
\begin{equation}
 \phi_i^{[n]} \; = \; A_{\{\nu_{k}(i)\}}\:\prod_{k=0}^{n}
 \beta_k^{\, m_k(i)}
 \label{eq:ansatz1}
\end{equation} 
where $m_{k}$ denotes the double-arrow potential in the $k$-fold
deflated tiling, and $\beta_{k}$ are $n+1$ free parameters.

\begin{figure}
 \centerline{\psfig{figure=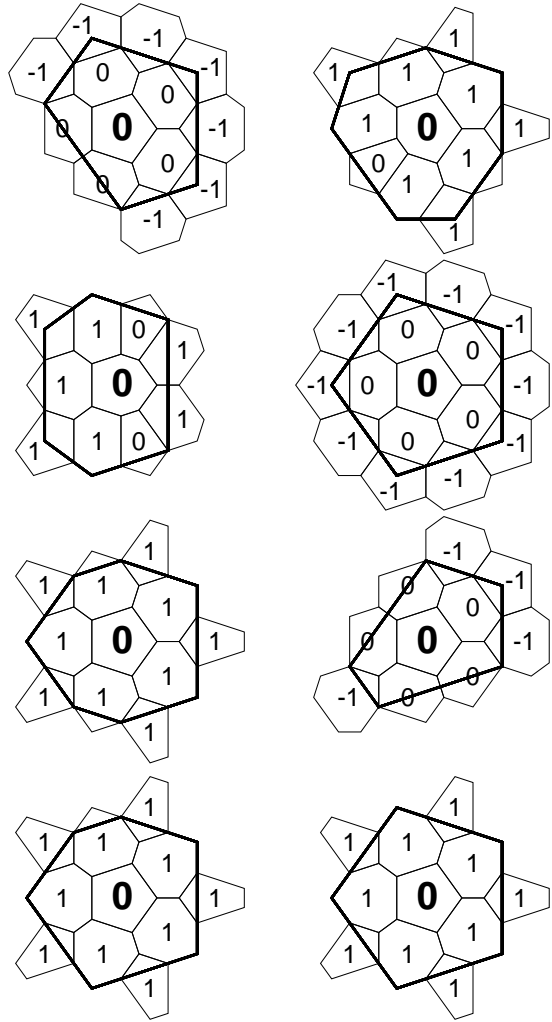,height=0.6\textheight}}
 \caption{Relation of the Voronoi cells of the Penrose tiling (thin
 lines) and their two-fold deflation (thick lines). The eight patches
 are obtained by a two-fold inflation of the eight vertex types (see
 Fig.~\protect\ref{fig:vertex_types}).  The numbers denote the change
 of the double-arrow potential with respect to the central Voronoi
 cell.\label{fig:vertex_types_inflated}}
\end{figure}

\begin{figure}
\centerline{\psfig{figure=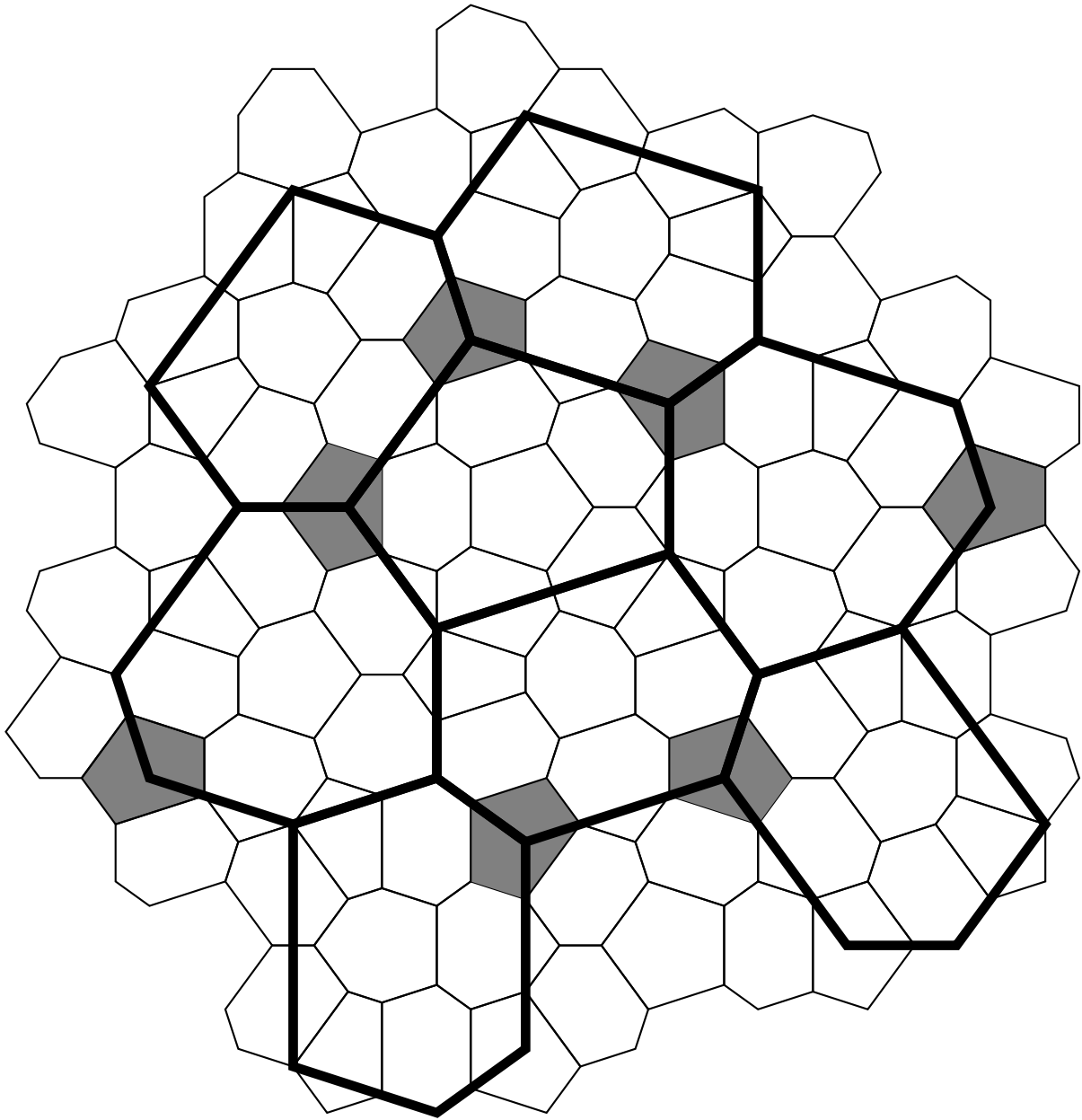,width=\columnwidth}}
\caption{Voronoi cells of a patch of the Penrose tiling (thin lines)
and of its two-fold deflation (thick lines). Shaded cells
corresponding to vertex type 1 cannot be uniquely assigned to a cell
of the deflated tiling.
\label{fig:ambiguous}}
\end{figure}

It is not completely obvious how to assign the vertex type
$\nu_{k}(i)$ of a site $i$ in the $k$-fold deflated tiling. Here, we
decided to use the concept of the Voronoi cell. We are looking for a
Voronoi cell of the deflated tiling that covers the Voronoi cell of
our site $i$ in the original tiling completely, or at least its
largest part. In Fig.~\ref{fig:vertex_types_inflated}, we show how the
Voronoi cells of the original and the two-fold deflated tiling relate
to each other. If a cell of the original tiling is shared between
several larger cells, we assign the vertex to the cell with the
maximum overlap.  However, there are still ambiguities when overlaps
of equal area occur. For instance, let us concentrate on the case
$n=2$. In the example shown in Fig.~\ref{fig:ambiguous}, one
recognizes that the cell corresponding to vertex type $1$ (cf.\
Fig.~\ref{fig:vertex_types}) may be dissected equally between the
cells corresponding to vertex types $2$ or $3$ of the deflated
tiling. In this case, we cannot assign the deflated vertex type
unequivocally.  Therefore, we demand that the corresponding terms in
the ansatz (\ref{eq:ansatz1}) are equal. In our example, this yields
the equation $A_{221}=A_{321}$ for the amplitudes
$A_{\nu_{2}\nu_{1}\nu_{0}}$ in the ansatz (\ref{eq:ansatz1}), labeled
by three digits according to the three vertex types. Considering also
the first deflation step, not shown in
Fig.~\ref{fig:vertex_types_inflated}, one finds another condition
$A_{222}=A_{232}$.

We now use the ansatz (\ref{eq:ansatz1}) to find solutions of the
tight-binding equations. Here, we restrict ourselves to the case
$n=2$. In order to set up the equations, we need to consider larger
patches which may be obtained by two-fold inflation of the $31$
second-order vertex types of Fig.~\ref{fig:vertex_types}.  Each of
these patches then leads to a number of equations.  Of the $8^3$
possible combinations of indices $\nu_{2},\nu_{1},\nu_{0}$, only $25$
occur in the Penrose tiling. Altogether we have to deal with a system
of $97$ equations in 32 variables, namely $25$ amplitudes
$A_{\nu_2\nu_1\nu_0}$, three variables $\beta_2$, $\beta_1$,
$\beta_0$, the four hopping parameters $d_1$, $d_2$, $d_3$, $d_4$, and
the energy $E$. We used {\em Mathematica}\/\cite{Wolfram} to solve
this system. As above, we find three sets of solutions, which we
express in terms of $\beta_{20}:=\beta_2\beta_0$ and $\beta_1$. They
have the following form.\newline 
{\em Solution (1'):} The $25$ amplitudes, without normalization, are
\begin{eqnarray}
& & A_{221} = A_{321} = -\beta_{20}\, (2\beta_{20}^{2}-\beta_{1}^2)
\,\beta_{0} \nonumber\\  
& & A_{222} = A_{322} = A_{232} = A_{532} = A_{732} = A_{832} = 
\beta_{20}^2\, \beta_{1}\,\beta_{0}\nonumber\\ 
& & A_{233} = A_{333} = \beta_{20}^2\, \beta_{0} \nonumber\\
& & A_{123} = A_{423} =  A_{623} = A_{133} = A_{633} = \beta_{20}^{3}
\nonumber\\ 
& & A_{654} = A_{174} = A_{484} = -\beta_{20}^2\, (2 \beta_{20}^2 - 
\beta_{1}^2)\nonumber\\
& & A_{365} = A_{217} = A_{548} = A_{748} = A_{848} =  
\beta_{20}\, \beta_{1}^{2}\,\beta_{0}^{2} \nonumber\\
& & A_{236} = A_{336} = -(2\beta_{20}^2 - \beta_{1}^2)\,\beta_{0}^{2} 
\label{eq:solution11}
\end{eqnarray}
and the transfer integrals and the energy are given by the same
expressions (\ref{eq:parameter1}) as for solution~(1), where now
\begin{equation}
b_{\pm} \; := \; \frac{\beta_{20}\,\beta_{1}}{\beta_{20}^{2}\pm\beta_{1}^{2}}
\; .\label{eq:bpmp}
\end{equation}
In contrast to the amplitudes, the transfer integrals and the energy
are hence expressed exclusively in terms of $\beta_{20}$ and
$\beta_{1}$, that is, the hopping parameters and the energy depend on
$\beta_2$ and $\beta_0$ only via the product $\beta_{20}$.\newline
{\em Solution (2'):} Here, the amplitudes read
\begin{eqnarray}
& & A_{221} = A_{321} = \beta_{20}^2\, \beta_{0}\nonumber\\ 
& & A_{222} = A_{322} = A_{232} = A_{532} = A_{732} = A_{832} = 
\beta_{20}\, \beta_{1}\, \beta_{0}\nonumber\\ 
& & A_{233} = A_{333} = \beta_{20}\, \beta_{0}\nonumber\\ 
& & A_{123} = A_{423} = A_{623} = A_{133} = A_{633} = \beta_{20}^{2}
\nonumber\\
& & A_{654} = A_{174} = A_{484} = \beta_{20}^{3}\nonumber\\
& & A_{365} = A_{217} = A_{548} = A_{748} = A_{848} = 
\beta_{1}^{2}\, \beta_{0}^{2}\nonumber\\
& & A_{236} = A_{336} = \beta_{20}\, \beta_{0}^{2} 
\label{eq:solution21}
\end{eqnarray}
and the parameters now follow from the expressions (\ref{eq:parameter2}) 
for solution~(2) with $b_{+}$ given by Eq.~(\ref{eq:bpmp}) and
\begin{equation}
\beta \; := \; \frac{\beta_{20}^{2}}{\beta_{1}^{2}} \; ;
\label{eq:beta}
\end{equation}
thus again they depend on $\beta_{0}$ and $\beta_{2}$ only via $\beta_{20}$.
\newline{\em Solution (3'):} Finally,
\begin{eqnarray}
& & A_{221} = A_{321} = \beta_{20}\, \beta_{1}^{2}\, \beta_{0}\nonumber\\ 
& & A_{222} = A_{322} = A_{232} = A_{532} = A_{732} = A_{832} = 
\beta_{20}^{2}\, \beta_{1}\, \beta_{0}\nonumber\\ 
& & A_{233} = A_{333} = \beta_{20}^{2}\, \beta_{0}\nonumber\\ 
& & A_{123} = A_{423} = A_{623} = A_{133} = A_{633} = 
\beta_{20}^{3}\nonumber\\
& & A_{654} = A_{174} = A_{484} = 
\beta_{20}^{2}\, \beta_{1}^{2}\nonumber\\
& & A_{365} = A_{217} = A_{548} = A_{748} = A_{848} = 
\beta_{20} \beta_{1}^{2}\, \beta_{0}^{2}\nonumber\\
& & A_{236} = A_{336} = \beta_{1}^{2}\, \beta_{0}^{2}
\label{eq:solution31}
\end{eqnarray}
where again the transfer integrals and the energy follow from the
previous expressions (\ref{eq:parameter3}) for solution~(3) by
replacing $b_{+}$ by Eq.~(\ref{eq:bpmp}) and $\beta$ by
Eq.~(\ref{eq:beta}).

These solutions comprise those found in the previous section. Indeed,
setting $\beta_2=\beta_1=1$ ($\beta_{20}=\beta_{0}$), we recover the
solutions (\ref{eq:solution1})--(\ref{eq:parameter3}), apart from a
common normalization factor $\beta_{0}^{2}$ in the amplitudes. In
addition, Eqs.~(\ref{eq:solution11})--(\ref{eq:solution31}) show that
the corresponding energy eigenvalues are infinitely degenerate. For
given values of $\beta_{20}$ and $\beta_{1}$, the Hamiltonian and the
energy $E$ are fixed, but the eigenfunctions still involve the free
parameter $\beta_{0}$. In other words, each choice of $\beta_{2}$ and
$\beta_{0}$ with the same product yields an eigenstate to the same
eigenvalue.  We note that infinite degeneracies in the spectrum were
previously observed in tight-binding models on the Penrose tilings.
One example is given by the confined degenerate states located at the
energy $E=0$ in the vertex model with $d_{1}\! =\! d_{2}\! =\! d_{3}\!
=\!  d_{4}\!  =\!  0$.\cite{ATFK,RS1} Also some of the critical,
self-similar eigenstates found in the center model appear to be
infinitely degenerate.\cite{TFA}

It is a question whether a larger number of deflation steps, i.e., a
larger value of $n$ in the ansatz (\ref{eq:ansatz1}), leads to further
solutions of the tight-binding equations. The larger $n$, the larger
is the number of sequences $\{\nu_{k}\}_{0\le k\le n}$ that occur, and
hence the number of independent amplitudes. Indeed, for $n=2$ we had
$25$ sequences, for $n=3$ and $n=4$ there are $49$ and $104$,
respectively. One might suspect that in the limit
$n\rightarrow\infty$, when the quantity of sequences tends to
infinity, every site is uniquely determined by its sequence, and hence
one should arrive at the complete solution in the limiting case.
However, this is not the case, which follows from the fact that ---
looking at it from the opposite point of view --- the dissection of a
cell under inflation may contain several copies of the same cell type.
Therefore, it is doubtful whether larger values of $n$ will lead to
new wave functions. For $n\le 4$, no solutions beyond
(\ref{eq:solution11})--(\ref{eq:solution31}) were
found. Nevertheless, this does not prove that further generalizations
might not be more rewarding.

\section{Multifractal Analysis}
\label{sec5}

Already a glimpse at Fig.~\ref{fig:wavefunctions} gives the impression
that the wave functions are self-similar. Let us therefore investigate
this property more thoroughly. To do this, we have to understand the
distribution of the double-arrow potential $m$ on the
tiling. Sutherland \cite{Suth} considered the transformation of single
and double arrows under two-fold inflation, and proved that the value
of the double-arrow potential changes at most by $2l$ under a
$2l$-fold inflation.

\begin{figure}
\centerline{\psfig{figure=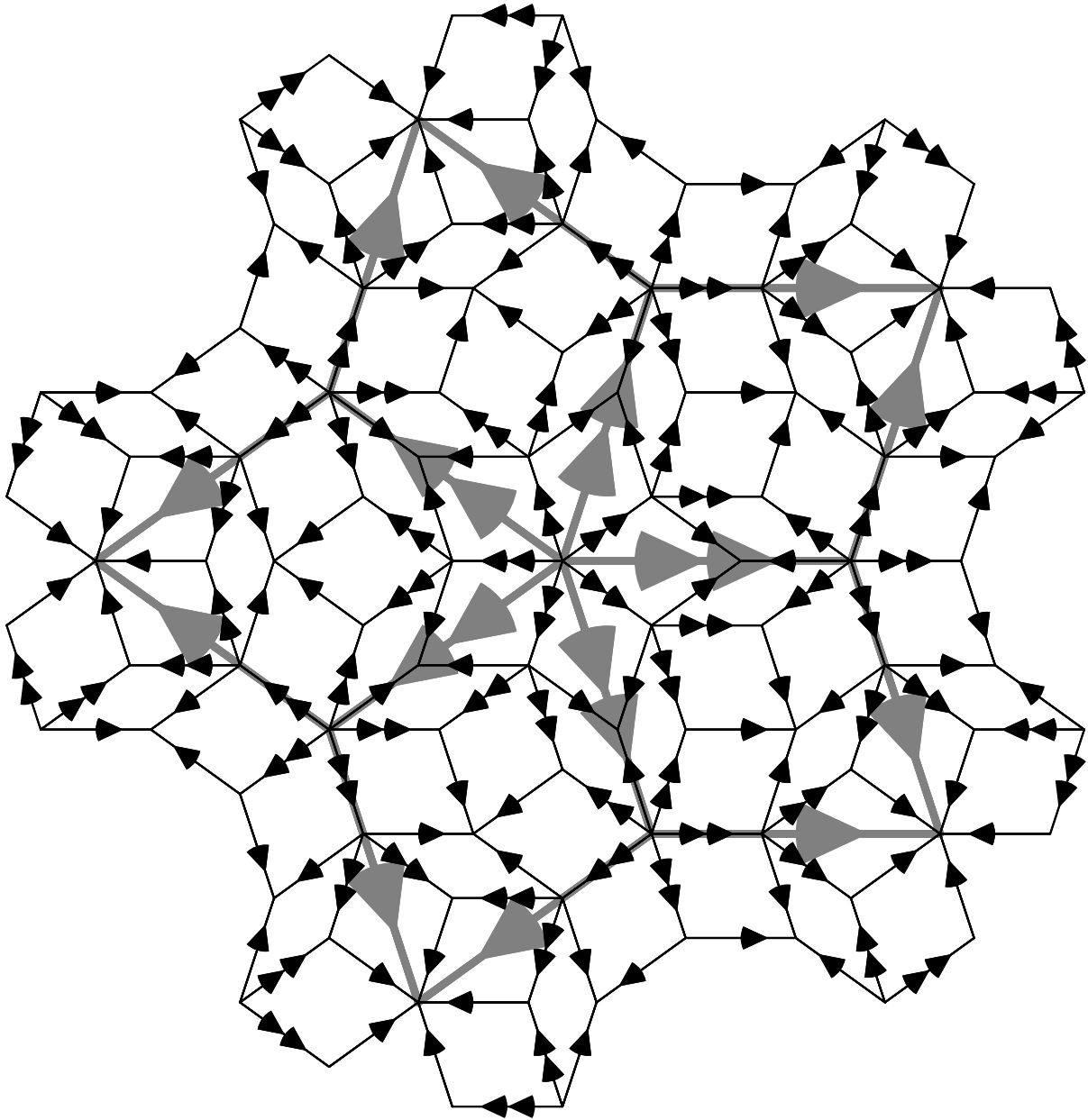,width=\columnwidth}}
\caption{A vertex of type 8 (grey) together
         with its two-fold inflation (black).\label{fig:vertex8}}
\end{figure}

For definiteness, let us consider a vertex of type 8 which has double
arrows pointing outwards in all five directions. In
Fig.~\ref{fig:vertex8}, we show this patch together with its two-fold
inflation. For the original patch, the values of the double-arrow
potential are $0$ at the center by our choice of normalization, and
$1$ elsewhere. In the inflated version, the potential takes values
between $0$ and $3$, compare Fig.~\ref{fig:pot1}. In what follows, we
use $2l$-fold inflations of this particular patch for the multifractal
analysis. In this case, the values of the double-arrow potential grow
linearly with the number of inflation steps. This may be different if
one starts from other initial patches, for example, starting from
vertex type $4$ results in a decreasing double-arrow potential,
corresponding to a different choice of the reference point for the
potential in the infinite tiling. We note that Refs.~\onlinecite{Suth}
and \onlinecite{TFA} used vertex type $4$, together with the opposite
direction of the arrows, which then also gives an increasing
potential.

Following Refs.~\onlinecite{TFA} and \onlinecite{Halsey}, we define a
partition function for the $2l$-fold inflated system
\begin{equation}
 \Gamma(q, \omega ; 2l) \;:=\; \frac{1}{N^{2q}_{}} \sum_i 
 \frac{|\phi_i|^{2q}}{{S_i}^{\omega / 2}} \; ,
 \label{eq:partition_function}
\end{equation}
where $N$ is the norm of the wave function on the finite patch, i.e.,
$N^2=\sum_i |\phi_i|^2$. Here, $S_i$ denotes the area of the Voronoi
cell of vertex $i$.  For a given $q$, there exists a certain number
$\omega(q)$ such that the partition function
(\ref{eq:partition_function}) is bounded (from above and below) in the
limit $l\rightarrow\infty$, i.e., it neither vanishes nor diverges.
The generalized dimensions
\begin{equation}
D_q\; :=\;\frac{\omega(q)}{q-1}
\label{eq:gendim}
\end{equation}
completely describe the multifractal properties of the wave function
$\phi$.

In an inflation step, the edge lengths of the rhombi are scaled by a
factor $\tau^{-1}$. Therefore, the area $S_\nu^{(2l)}$ of a Voronoi
cell corresponding to vertex type $\nu$ of the $2l$-fold inflated tiling
is given by
\begin{equation} 
S_{\nu}^{(2l)} \;=\; \tau^{-4l} S_\nu^{(0)} \; .
\label{eq:Voronoi_scale} 
\end{equation}
For simplicity, we restrict our analysis to the ansatz
(\ref{eq:ansatz}) for the wave function. In fact, since only the
absolute values of the wave function amplitudes enter in
Eq.~(\ref{eq:partition_function}), this also applies to the solutions
(\ref{eq:gensol}).  Substituting the ansatz into
Eq.~(\ref{eq:partition_function}) yields
\begin{eqnarray}
 \Gamma(q, \omega ; 2l) & = & 
\frac{1}{N^{2q}_{}} \sum_i 
\frac{|A_{\nu(i)}\beta^{\, m(i)}|^{2q}}{\tau^{-2l\omega}
\bigl(S_{\nu(i)}^{(0)}\bigr)^{\frac{\omega}{2}}}
\nonumber \\
& = & 
\sum_{\nu=1}^{8} \biggl[
\frac{|A_\nu|^{2q}\,\tau^{2l\omega}}
{\bigl(S_{\nu}^{(0)}\bigr)^{\frac{\omega}{2}}N^{2q}_{}} \!
\sum_{m=0}^{2l} |\beta|^{2qm}\, V_\nu(m;2l) \biggr]
 \label{eq:partition_function1}
\end{eqnarray}
where $V_{\nu}(m;2l)$ denotes the number of vertices of type $\nu$
with potential $m$ multiplied by the area of the corresponding Voronoi
cell after $2l$ inflation steps. 

In order to calculate $V_{\nu}(m;2l)$, we consider the transformation
of the Voronoi cells of the eight vertex types under a two-fold
inflation, compare Figs.~\ref{fig:vertex_types} and
\ref{fig:vertex_types_inflated}.  {}From this, one derives recursion
relations for the distributions $V_{\nu}(m;2l)$ by counting the number
of inflated cells that are covered by the original cell. For example,
as shown in the lower right corner of
Fig.~\ref{fig:vertex_types_inflated}, the Voronoi cell corresponding
to the vertex type $8$ with a potential $m$ turns into: (i) one cell
of type $8$ with potential $m$; (ii) five cells of type $2$ with
potential $m+1$; and (iii) five fractional parts (each with an area
fraction of $(4-\tau)/11\approx 0.216542$) of type-$6$ cells with
potential $m+1$. Conversely, a cell of type $8$ in the inflated patch
may stem from a vertex of type $5$, $7$, or $8$, each of those
yielding precisely one complete cell of type $8$.  Considering all
vertex types, and computing the fractional areas involved, one arrives
at recursion relations
\begin{equation}
V_{\nu}(m;2l+2) \; = \;
\sum_{\sigma=-1}^{1} \sum_{\mu=1}^{8} M^{(\sigma)}_{\nu,\mu}\, 
V_{\mu}(m+\sigma;2l)
\label{eq:recursion}
\end{equation}
with three $8\times 8$ matrices $M^{(-1)}$, $M^{(0)}$, and
$M^{(1)}$. The quantities we need are certain transforms
$\tilde{V}_{\nu}(\beta ;2l)$ of $V_{\nu}(m;2l)$, defined as
\begin{equation}
\tilde{V}_{\nu}(\beta ;2l) \;:=\; \sum_{m=0}^{2l} \beta^m V_{\nu}(m;2l) 
\end{equation}
see Eq.~(\ref{eq:partition_function1}). {}From the recursion relations
(\ref{eq:recursion}), one finds that the transforms
$\tilde{V}_{\nu}(\beta ;2l)$ for two successive inflation steps are
related by
\begin{equation}
\tilde{V}_{\nu}(\beta ; 2l + 2) \;=\; \sum_{\mu=1}^{8}
\tilde{M}_{\nu,\mu}(\beta)\, \tilde{V}_{\mu}(\beta ; 2l) \; .
\end{equation}
where  the matrix $\tilde{M}(\beta)$ reads as follows
\widetext
\begin{equation}
\tilde{M}(\beta) = 
\renewcommand{\arraystretch}{1.5}
\left( 
\begin{array}{@{}cccccccc@{}}
\scriptstyle{\frac{21+6\tau}{59}\beta} & 
\scriptstyle{\frac{52-2\tau}{59}\beta^{-1}} & 
\scriptstyle{\frac{52-2\tau}{59}\beta^{-1}} & 
\scriptstyle{\frac{35+10\tau}{59}\beta} & 
\scriptstyle{0} & 
\scriptstyle{\frac{7+2\tau}{59}\beta} & 
\scriptstyle{0} & 
\scriptstyle{0} \\
\scriptstyle{\frac{-16+20\tau}{29}\beta} & 
\scriptstyle{\frac{95-10\tau}{29}\beta^{-1}} & 
\scriptstyle{\frac{33-5\tau}{29}\beta^{-1}} & 
\scriptstyle{\frac{-20+25\tau}{29}\beta} & 
\scriptstyle{\frac{153-10\tau}{29}\beta^{-1}} & 
\scriptstyle{\frac{-12+15\tau}{29}\beta}& 
\scriptstyle{\frac{149-5\tau}{29}\beta^{-1}} & 
\scriptstyle{5\beta^{-1}} \\
\scriptstyle{\frac{140-6\tau}{31}} & 
\scriptstyle{2\beta^{-1}} & 
\scriptstyle{\frac{77+6\tau}{31}\beta^{-1}} & 
\scriptstyle{5} & 
\scriptstyle{0} & 
\scriptstyle{\frac{125-12\tau}{31}} & 
\scriptstyle{0} & 
\scriptstyle{0} \\
\scriptstyle{1} & 
\scriptstyle{0} & 
\scriptstyle{0} & 
\scriptstyle{1} & 
\scriptstyle{0} & 
\scriptstyle{1} & 
\scriptstyle{0} & 
\scriptstyle{0} \\
\scriptstyle{0} & 
\scriptstyle{0} & 
\scriptstyle{1} & 
\scriptstyle{0} & 
\scriptstyle{0} & 
\scriptstyle{0} & 
\scriptstyle{0} & 
\scriptstyle{0} \\
\scriptstyle{0} & 
\scriptstyle{\frac{7+\tau}{11}+\frac{8-2\tau}{11}\beta^{-1}} & 
\scriptstyle{\frac{14+2\tau}{11}} & 
\scriptstyle{0} & 
\scriptstyle{\frac{20-5\tau}{11}\beta^{-1}} & 
\scriptstyle{0} & 
\scriptstyle{\frac{20-5\tau}{11}\beta^{-1}} & 
\scriptstyle{\frac{20-5\tau}{11}\beta^{-1}} \\
\scriptstyle{0} & 
\scriptstyle{1} & 
\scriptstyle{0} & 
\scriptstyle{0} & 
\scriptstyle{0} & 
\scriptstyle{0} & 
\scriptstyle{0} & 
\scriptstyle{0} \\
\scriptstyle{0} & 
\scriptstyle{0} & 
\scriptstyle{0} & 
\scriptstyle{0} & 
\scriptstyle{1} & 
\scriptstyle{0} & 
\scriptstyle{1} & 
\scriptstyle{1} \end{array}
\right)\label{eq:matrix}
\renewcommand{\arraystretch}{1}
\end{equation}
\narrowtext compare Ref.~\onlinecite{Suth}. It is related to the
matrices $M^{(\sigma)}$ (\ref{eq:recursion}) by
\begin{equation}
\tilde{M}(\beta) \; =\; \sum_{\sigma=-1}^{1} \beta^{-\sigma}\, M^{(\sigma)}
\end{equation}
hence the elements of $M^{(\sigma)}$ are nothing but the coefficients of 
$\beta^{-\sigma}$ of the elements of $\tilde{M}(\beta)$.

\begin{figure}
\centerline{\psfig{figure=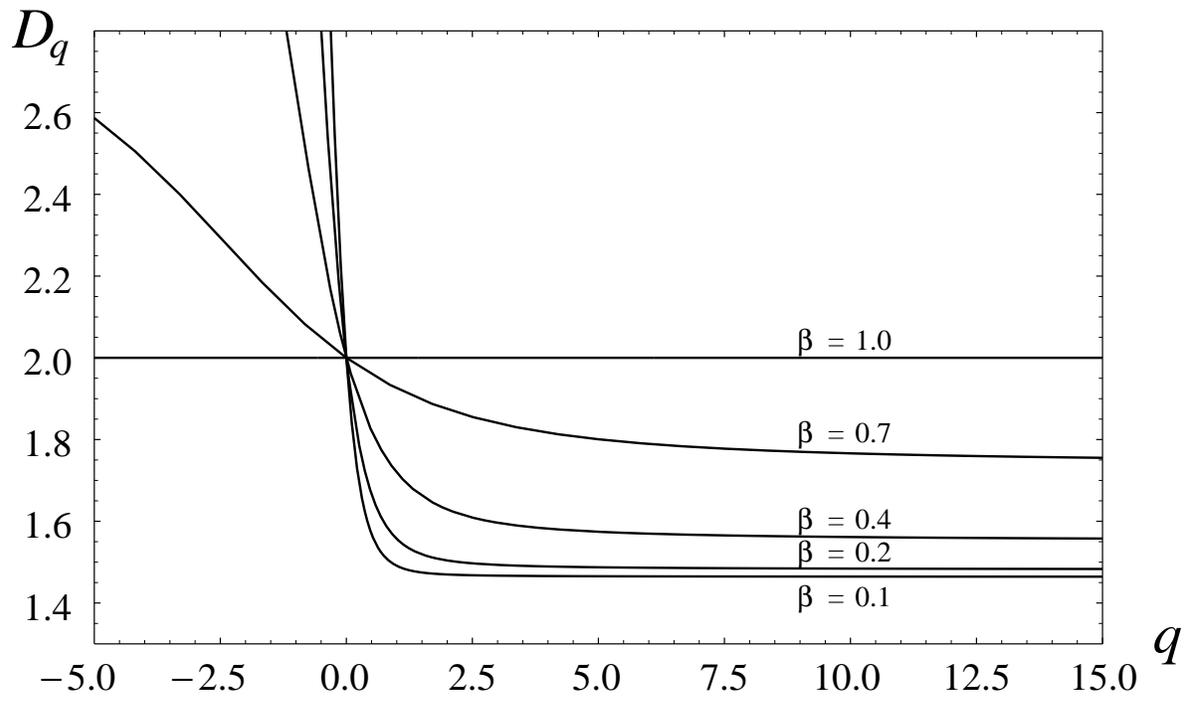,width=\columnwidth}}
\caption{The generalized dimensions $D_q$ for several
         values of $\beta$.\label{fig:fract_exponent}}
\end{figure}

The asymptotic behavior (for $l\rightarrow\infty$) of
$\tilde{V}(\beta ;2l)$ is governed by the eigenvalue $\Omega(\beta)$ of
$\tilde{M}(\beta)$ with largest modulus
\begin{equation}
\tilde{V}_{\nu}(\beta ;2l) \;\sim\;
\Omega^{l}(\beta) f_{\nu}(\beta) 
\end{equation}
where $f(\beta)$ is the corresponding eigenvector.  For positive
values of $\beta$, the largest eigenvalue (in absolute value) is
positive and unique, because the third power of $\tilde{M}(\beta)$ is a
positive matrix. Calculating the norm
\begin{eqnarray}
N^2 &=& \sum_{i} |\phi_i|^2 \;=\;
\sum_{\nu=1}^{8} |A_{\nu}|^2 \tilde{V}_{\nu}(|\beta|^{2} ;2l)\nonumber\\
&\sim& \Omega^{l}(|\beta|^{2})
\end{eqnarray} 
and substituting the asymptotic behavior of $\tilde{V}_{\nu}(\beta ;
2l)$ into (\ref{eq:partition_function1})
\begin{eqnarray}
\Gamma(q, \omega ; 2l) & = & 
\sum_{\nu=1}^{8}
\frac{|A_\nu|^{2q}\,\tau^{2l\omega}}
{\bigl(S_{\nu}^{(0)}\bigr)^{\omega /2}N^{2q}_{}}
\tilde{V}_{\nu}(|\beta|^{2q} ;2l)\nonumber\\
&\sim& \left[\frac{\tau^{2\omega}\Omega(|\beta|^{2q})}
{\Omega^{q}(|\beta|^{2})}\right]^{l}
\end{eqnarray}
leads us to the conclusion that the partition function 
$\Gamma(q,\omega ; 2l)$ can be bounded if and only if 
\begin{equation}
  \omega(q) \;=\; \frac{1}{2\log{\tau}}\;\log{\left[
\frac{\Omega^{q}(|\beta|^2)}
     {\Omega(|\beta|^{2q})}\right]} \; .
\end{equation}
In Fig.~\ref{fig:fract_exponent}, we present the fractal exponent
$D_q$ (\ref{eq:gendim}) for several values of $\beta$.  For $\beta=1$,
the wave function does not depend on the potential $m$ and takes at
most four different values according to the translation class of the
site.  In this case $D_q$ is constant. The smaller $|\beta|$, the
faster the wave function decays, leading to a steeper curve $D_q$ as a
function of $q$.

Concerning the matrix $\tilde{M}(\beta)$ (\ref{eq:matrix}), we remark
that its eigenvalues and eigenvectors are connected to the frequencies
of the vertex types, which count how often a certain vertex type
occurs in the Penrose tiling.  Indeed, if we set $\beta=1$, we obtain
a substitution matrix for the inflation rules in the Penrose tiling.
Therefore, according to the Perron-Frobenius theorem, the eigenvector
$f(1)$ corresponding to the the eigenvalue with largest modulus
$\Omega(1)$ should reproduce the relative frequencies of the vertex
types in the tiling.  We calculated numerically $f(1)$ and found
perfect agreement with the known frequencies.\cite{Jaric,BKSZ}

We note that the multifractal analysis can be carried out for the
generalized eigenstates (\ref{eq:solution11})--(\ref{eq:solution31})
analogously.  However, it becomes more complicated because we have to
consider the substitution matrix of vertices labeled by the inflated
vertex types, which results in a $25\times 25$ matrix.

\section{Conclusions}
\label{sec6}

We constructed exact non-normalizable eigenfunctions for certain
vertex-type tight-binding models on the rhombic Penrose tiling. The
construction is based on a potential $m$ derived from the matching
rules of the Penrose tiling that had been introduced in a similar
context previously.\cite{Suth,TFA} We consider several generalizations
of the ansatz for the wave functions, which show that the eigenstates
we found are infinitely degenerate. Further generalizations can be
investigated in a systematic way, and may lead to a wider class of
accessible wave functions. We hope to report on this, and on the
application of this ansatz to other quasiperiodic tight-binding models
(particularly for the three-dimensional case) in the future.

{}From the ansatz, it is apparent that the eigenfunctions
(\ref{eq:solution1}), (\ref{eq:solution2}) and (\ref{eq:solution3})
reflect the distribution of the potential $m$ on the lattice. The
multifractal analysis of the eigenstates therefore reduces to the
analysis of the distribution of the potential which was already
considered by Sutherland.\cite{Suth} It shows that the wave functions
are critical, i.e., neither extended nor exponentially localized, as
typically expected in two-dimensional quasiperiodic tight-binding
models.

The present work is a generalization of the ideas of
Refs.~\onlinecite{Suth} and \onlinecite{TFA}, and we recover the
solutions found by Sutherland\cite{Suth} as a special case. In
Ref.~\onlinecite{TFA}, the authors considered a center model on the
Penrose tiling, where they also found infinitely degenerate critical
eigenstates. It is interesting to note that all exactly known
eigenstates in such models, including the confined
states,\cite{KS,ATFK,FATK} appear at energies with infinite
degeneracy.  At present, we do not know whether there is a deeper
reason for this observation.

\acknowledgements The authors thank M.~Baake for discussions and
helpful comments. Financial support from DFG is gratefully
acknowledged.

\clearpage
\narrowtext

\clearpage
\narrowtext

\clearpage
\narrowtext

\clearpage
\widetext

\clearpage
\narrowtext

\clearpage
\widetext

\clearpage
\narrowtext

\clearpage
\narrowtext

\clearpage
\narrowtext

\clearpage
\narrowtext

\end{document}